\documentclass[12pt,preprint]{aastex}

\shorttitle{Cosmic-ray Sidereal Anisotropy}
\shortauthors{Amenomori et al.}

\begin{document}


\title{Large-Scale Sidereal Anisotropy of Galactic Cosmic-Ray 
Intensity Observed by the Tibet Air Shower Array}
\author{
M.~Amenomori\altaffilmark{1},
S.~Ayabe\altaffilmark{2},
S.W.~Cui\altaffilmark{3},
Danzengluobu\altaffilmark{4},
L.K.~Ding\altaffilmark{3},
X.H.~Ding\altaffilmark{4},
C.F.~Feng\altaffilmark{5},
Z.Y.~Feng\altaffilmark{6},
X.Y.~Gao\altaffilmark{7},
Q.X.~Geng\altaffilmark{7},
H.W.~Guo\altaffilmark{4},
H.H.~He\altaffilmark{3},
M.~He\altaffilmark{5},
K.~Hibino\altaffilmark{8},
N.~Hotta\altaffilmark{9},
Haibing~Hu\altaffilmark{4},
H.B.~Hu\altaffilmark{3},
J.~Huang\altaffilmark{10},
Q.~Huang\altaffilmark{6},
H.Y.~Jia\altaffilmark{6},
F.~Kajino\altaffilmark{11},
K.~Kasahara\altaffilmark{12},
Y.~Katayose\altaffilmark{13},
C.~Kato\altaffilmark{14},
K.~Kawata\altaffilmark{10},
Labaciren\altaffilmark{4},
G.M.~Le\altaffilmark{15},
J.Y.~Li\altaffilmark{5},
H.~Lu\altaffilmark{3},
S.L.~Lu\altaffilmark{3},
X.R.~Meng\altaffilmark{4},
K.~Mizutani\altaffilmark{2},
J.~Mu\altaffilmark{7},
K.~Munakata\altaffilmark{14},
A.~Nagai\altaffilmark{16},
H.~Nanjo\altaffilmark{1},
M.~Nishizawa\altaffilmark{17},
M.~Ohnishi\altaffilmark{10},
I.~Ohta\altaffilmark{9},
H.~Onuma\altaffilmark{2},
T.~Ouchi\altaffilmark{8},
S.~Ozawa\altaffilmark{10},
J.R.~Ren\altaffilmark{3},
T.~Saito\altaffilmark{18},
M.~Sakata\altaffilmark{11},
T.~Sasaki\altaffilmark{8},
M.~Shibata\altaffilmark{13},
A.~Shiomi\altaffilmark{10},
T.~Shirai\altaffilmark{8},
H.~Sugimoto\altaffilmark{19},
M.~Takita\altaffilmark{10},
Y.H.~Tan\altaffilmark{3},
N.~Tateyama\altaffilmark{8},
S.~Torii\altaffilmark{8},
H.~Tsuchiya\altaffilmark{10},
S.~Udo\altaffilmark{10},
T.~Utsugi\altaffilmark{8},
B.S.~Wang\altaffilmark{3},
H.~Wang\altaffilmark{3},
X.~Wang\altaffilmark{2},
Y.G.~Wang\altaffilmark{5},
H.R.~Wu\altaffilmark{3},
L.~Xue\altaffilmark{5},
Y.~Yamamoto\altaffilmark{11},
C.T.~Yan\altaffilmark{3},
X.C.~Yang\altaffilmark{7},
S.~Yasue\altaffilmark{14},
Z.H.~Ye\altaffilmark{15},
G.C.~Yu\altaffilmark{6},
A.F.~Yuan\altaffilmark{4},
T.~Yuda\altaffilmark{10},
H.M.~Zhang\altaffilmark{3},
J.L.~Zhang\altaffilmark{3},
N.J.~Zhang\altaffilmark{5},
X.Y.~Zhang\altaffilmark{5},
Y.~Zhang\altaffilmark{3},
Yi~Zhang\altaffilmark{3},
Zhaxisangzhu\altaffilmark{4}, and
X.X.~Zhou\altaffilmark{6} \\
(The Tibet AS$\gamma$ Collaboration)}

\altaffiltext{1}{ Department of Physics, Hirosaki University, Hirosaki 036-8561, Japan}
\altaffiltext{2}{ Department of Physics, Saitama University, Saitama 338-8570, Japan }
\altaffiltext{3}{ Institute of High Energy Physics, Chinese Academy of Sciences, Beijing 100039, China }
\altaffiltext{4}{ Department of Mathematics and Physics, Tibet University, Lhasa 850000, China }
\altaffiltext{5}{  Department of Physics, Shandong University, Jinan 250100, China }
\altaffiltext{6}{ Institute of Modern Physics, South West Jiaotong University, Chengdu 610031, China }
\altaffiltext{7}{ Department of Physics, Yunnan University, Kunming 650091, China }
\altaffiltext{8}{  Faculty of Engineering, Kanagawa University, Yokohama 221-8686, Japan}
\altaffiltext{9}{ Faculty of Education, Utsunomiya University, Utsunomiya 321-8505, Japan}
\altaffiltext{10}{Institute for Cosmic Ray Research, University of Tokyo, Kashiwa 277-8582, Japan }
\altaffiltext{11}{Department of Physics, Konan University, Kobe 658-8501, Japan}
\altaffiltext{12}{ Faculty of Systems Engineering, Shibaura Institute of Technology, Saitama 330-8570, Japan}
\altaffiltext{13}{ Faculty of Engineering, Yokohama National University, Yokohama 240-8501, Japan }
\altaffiltext{14}{ Department of Physics, Shinshu University, Matsumoto 390-8621, Japan}
\altaffiltext{15}{ Center of Space Science and Application Research, Chinese Academy of Sciences, Beijing 100080, China }
\altaffiltext{16}{ Advanced Media Network Center, Utsunomiya University, Utsunomiya 321-8585, Japan}
\altaffiltext{17}{ National Institute for Informatics, Tokyo 101-8430, Japan}
\altaffiltext{18}{ Tokyo Metropolitan College of Aeronautical Engineering, Tokyo 116-0003, Japan}
\altaffiltext{19}{ Shonan Institute of Technology, Fujisawa 251-8511, Japan}

\begin{abstract}
We present the large-scale sidereal anisotropy of
galactic cosmic-ray intensity in the multi-TeV region observed with the Tibet-III
air shower array during the period from 1999 through 2003. 
The sidereal daily variation of cosmic rays
observed in this experiment shows an excess of relative intensity around
$4\sim7 $ hours local sidereal time, as well as a deficit around 12 hours local
sidereal time.
While the amplitude of the excess is not significant when averaged over all 
declinations, the excess in individual declinaton bands becomes larger and
clearer as the viewing direction moves toward the south. The
maximum phase of the excess intensity changes from $\sim$7 at the northern
hemisphere to $\sim$4 hours at the equatorial region. We also show that both 
the amplitude and the phase of the first harmonic vector of the daily variation 
are remarkably independent of primary energy in the multi-TeV region. 
This is the first result determining the energy and declination dependences 
of the full 24-hour profiles of the sidereal daily variation in the multi-TeV 
region with a single air shower experiment.
\end{abstract}

\keywords{cosmic rays ---  diffusion --- ISM:magnetic field}

\section{Introduction}
The directional anisotropy of galactic cosmic-ray intensity in the multi-TeV region 
gives us an important piece of information about the magnetic structure of the heliosphere 
and/or the local interstellar space surrounding the heliosphere, through which cosmic rays 
propagate to the Earth. 
The galactic anisotropy has been measured via the sidereal daily variation (SDV) 
of cosmic-ray intensity recorded in a fixed directional channel of the detector on the 
spinning Earth. 
On the basis of the SDV observed in the $1-100$ TeV region, most of the previous investigations 
reported on first harmonic vector with small amplitude ($1-10\times10^{-4}$ or $0.01-0.1$\%) 
and a phase of maximum somewhere between $23-3$ hours in the local sidereal time 
(LST)~\citep{jack1966,cutler1981,AN1985,jack1986,nagashima-1989,Bergamasco,cg1991,EASTOP,mk-kamioka}.
Based on a harmonic analysis, these observations are consistent with the large-scale diffusive 
propagation of cosmic rays in the Galaxy~\citep{shibatat}, but there is no consensus currently  
for a production mechanism in the local interstellar region surrounding the heliosphere 
that would reproduce the full 24-hour profile of the SDV ~\citep{jack1966,AN1985,nagashima-1989,
Bergamasco}. 
This is partly due to the lack of the information on the dependence of the anisotropy on
declination. 
Also the first harmonic vector is not always sufficient to precisely represent the full 24-hour 
profile of the SDV, since the SDV often has significant higher-order harmonics, including but
not limited to a second harmonic. 
Both the 24-hour profile of the SDV and its declination dependence can only be  obtained by 
utilizing multiple high-count observations of the celestial sphere which has not been possible 
until recently.

A continuous observation of the 24-hour profile of the SDV over 12 years with 
an air-shower (AS) detector at Mt. Norikura in Japan revealed that the 
SDV of ~10 TeV cosmic-ray intensity exhibits a deficit with a minimum around 12 hours 
LST~\citep{nagashima-1989}. 
Similar profiles has been reported by several other experiments in the same energy 
region~\citep{AN1985,Bergamasco,EASTOP,mk-kamioka}. 
The intensity deficit has also been seen in the sub-TeV region, below 1 TeV covered by  
underground muon detectors~\citep*[hereafter referred to as NFJ]{nagashima-1998}.
In addition to this anisotropy,  which they named ``Galactic'' anisotropy, NFJ has also found a 
new anisotropy component causing an excess of intensity with a maximum around 6 hours LST 
in the sub-TeV region. 
The amplitudes of both of these two anisotropy components increase with increasing energy 
up to $\sim1$ TeV, as higher energy particles become less sensitive to the solar modulation 
effects masking the galactic anisotropy. 
An analysis of the SDVs recorded in a total of 48 directional channels of the underground 
muon detectors monitoring both the northern and southern sky  
indicated that the maximum phase of the new anisotropy component,   
which is $\sim6$ hours in the northern hemisphere, shifts toward earlier times as the 
declination of the incident cosmic-rays moves south toward the equator~\citep{thn-1998,thn-1999}.
Since this new anisotropy component was not seen by the omnidirectional measurement
of the Norikura AS array at $\sim10$ TeV, which was not capable of resolving the declination 
of the incident direction, NFJ suggested that the amplitude of this anisotropy 
decreases with increasing energy between 1 and 10 TeV. 
As such an energy spectrum is consistent with an anisotropy due to the possible acceleration of
particles arriving from the heliotail direction, i.e., $\sim6$ hours in LST, this new component 
has been named ``Tail-In'' anisotropy~(NFJ).

A two-dimensional map of the SDV obtained with the Tibet III air shower array has been
published elsewhere~\citep{tibet-all-sky,Wu_talk}.
In this paper, we present an analysis of the SDV observed by the 
Tibet III air shower array, examining the precise form of both the full 
24-hour profile and the energy and declination dependences of the SDV with high 
statistics and a good angular resolution of the incident direction of primary particles. 
The reliability of both the measurement and the analysis in the Tibet III experiment 
in the multi-TeV region are assured by the successful observation of the Compton-Getting (CG) 
anisotropy due to the orbital motion of the Earth around the Sun~\citep[hereafter referred to 
as paper 1]{tibet-cg}.

\section{Experiment}
The Tibet air shower experiment has been successfully operating at Yangbajing
($90.522^{\circ}$E, $30.102^{\circ}$N, 4300 m above sea level) in Tibet, China since 1990.
The array, originally  constructed in 1990, was gradually upgraded by increasing the number
of counters~\citep{tibet2000,tibet2002}. The Tibet III array, used in the present analysis,
was completed in the late fall of 1999. This array consists of 533
scintillation counters of $0.5 \mathrm{m^2}$ each placed on a $7.5 \mathrm{m}$ 
square grid with an enclosed area of $22,050 \mathrm{m^2}$ and each viewed by a fast-timing (FT)
photomultiplier tube. A $0.5 \mathrm{cm}$ thick lead plate is placed on the top of each
counter in order to increase the array sensitivity by converting $\gamma$-rays
into electron-positron pairs.

An event trigger signal is issued when any fourfold coincidence occurs in the
FT counters recording more than 0.6 particles, resulting in a trigger rate
of about 680 Hz at a few-TeV threshold energy. We collected $5.4 \times 10^{10}$
events by the Tibet III array during 918 live days from 1999 November to
2003 November. After some simple data selections (software trigger condition
of any fourfold coincidence in the FT counters recording more than 0.8
particles in charge, zenith angle of arrival direction $< 45^{\circ}$, air shower
core position located in the array, etc), $3.0 \times 10^{10}$ events remained for
further analysis. The declination of incident direction of these events ranges
from $-15^{\circ}$ and $75^{\circ}$.

The performance of the Tibet III array is also examined by employing a full
Monte Carlo (MC) simulation in the energy range from 0.3 to 1000 TeV. We use
the CORSIKA version 6.004 code~\citep{corsika} and the QGSJET model~\citep{qgsjet} 
for the generation
of air shower events and the EPICS UV8.00 code~\footnote{http://cosmos.n.kanagawa-u.ac.jp/EPICSHome/}
of shower particles with scintillation counters. Primary cosmic-ray particles
are sampled from the energy spectrum made by a compilation of direct
observational data. The primary cosmic-ray energy is estimated by $\sum \rho_{\mathrm{FT}}$
which is the sum of the number of particles/$\mathrm{m^2}$ for each FT counter.
According to the result of the simulation, $\sum \rho_{\mathrm{FT}}=100$
corresponds approximately to 10 TeV primary cosmic-ray energy~\citep{tibet2003-mrk421}.



\section{Analysis}
The selected air shower events are subsequently histogrammed into hourly bins in
LST (366 cycle/year), according to the time, incident direction, and air
shower size of each event. 
In order to check the seasonal change in the daily variation, we constructed the 
histogram for each month and corrected it for the observation live time varying 
month to month. 
Following paper 1, we first
obtain the daily variation for each of East and West (E- and W-) incident
events and adopt the ``East-West''(E-W) subtraction method to eliminate
meteorological effects and possible detector biases. 
By dividing the difference by the hour angle separation between the mean E- and 
W-incident directions averaged over the E- and W-incident events, we obtain the 
``differential'' variation ``$D(t)$'' in sidereal time. The physical variation 
``$R(t)$'', which is expected to be free from spurious effects, can be reconstructed by
integrating $D(t)$ in sidereal time $t$. 
Hereafter, we make statistical arguments on the basis of the differential
variation $D(t)$ to avoid the difficulty in estimating the error in $R(t)$. 
By adopting this method, the spurious variation contained in the average
daily variation is reduced to $\leq 0.01$\%, which is less than 20\% of the CG
anisotropy with the amplitude of $\sim 0.05$\% in the solar time. 
A spurious variation in sidereal time also could be produced from a seasonal change 
of the daily variation in solar time. We confirmed, however, that the variations with 
non-physical frequencies, i.e., the variations in the anti-sidereal time (364 c/y) 
(also called the psuedo-sidereal time) and in the extended-sidereal time (367 c/y), 
are negligibly small. 
This ensures that the seasonal changes of the solar and sidereal daily variation are 
both negligible. For more details of the method, readers are referred to paper 1 
and~\citet{nagashima-1989}.

The data are divided into four data samples according to representative primary 
energies of 4.0, 6.2, 12, and 53 TeV. Each of these representative energies is 
calculated as the mode value of the logarithmic energy of each event produced by 
the MC simulation.

In our analysis, we first calculate the harmonic vector by best-fitting the 
``differential variation'' $D(t)$ with the harmonic function
\begin{equation}
D^{cal}(t) = A_{D}\cos{\frac{2\pi}{24}(t-\Phi_{D})},
\label{eq:1st}
\end{equation}
where $A_{D}$ and $\Phi_{D}$ are the amplitude and phase of the first harmonic vector
of $D(t)$ and $t$ is local sidereal time in hours. By integrating equation (\ref{eq:1st}) 
in time $t$ , we arrive at the amplitude and phase of the physical variation $R(t)$, as
\begin{equation}
A_{R} = \frac{24}{2\pi}A_{D}, \quad\quad \Phi_{R} = \Phi_{D} + 6. 
\end{equation}
The errors of $A_{R}$ and $\Phi_{R}$ are deduced from errors of the observed $D(t)$.

\section{Results and Discussions}
\begin{figure}
\includegraphics[scale=0.6,angle=270]{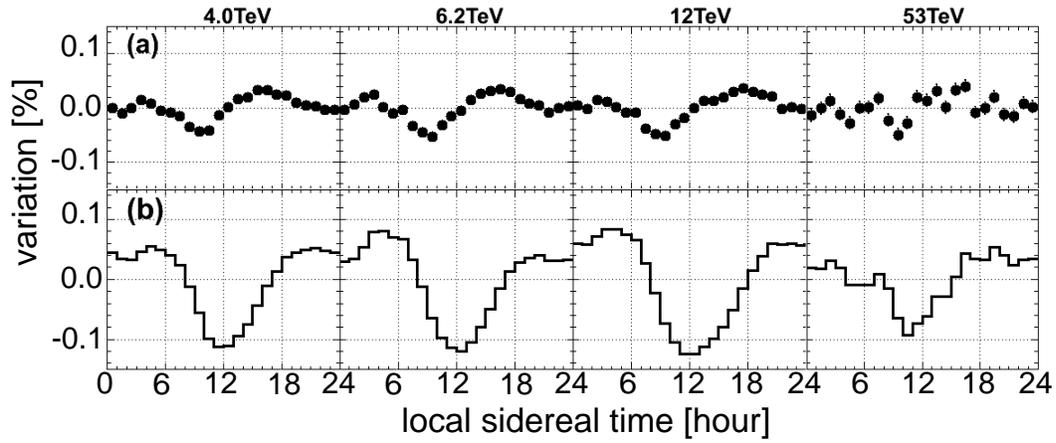}
\caption{The sidereal daily variations averaged over all declinations as a function of 
representative energies of 4.0, 6.2, 12, and 53 TeV. 
The upper panels (a) show the differential variations $D(t)$'s, 
while the lower panels (b) display the physical variations $R(t)$'s. 
The error bars are statistical.
\label{sit-D_R_avedir}
}
\end{figure}

\begin{figure}
\epsscale{0.8}
\plotone{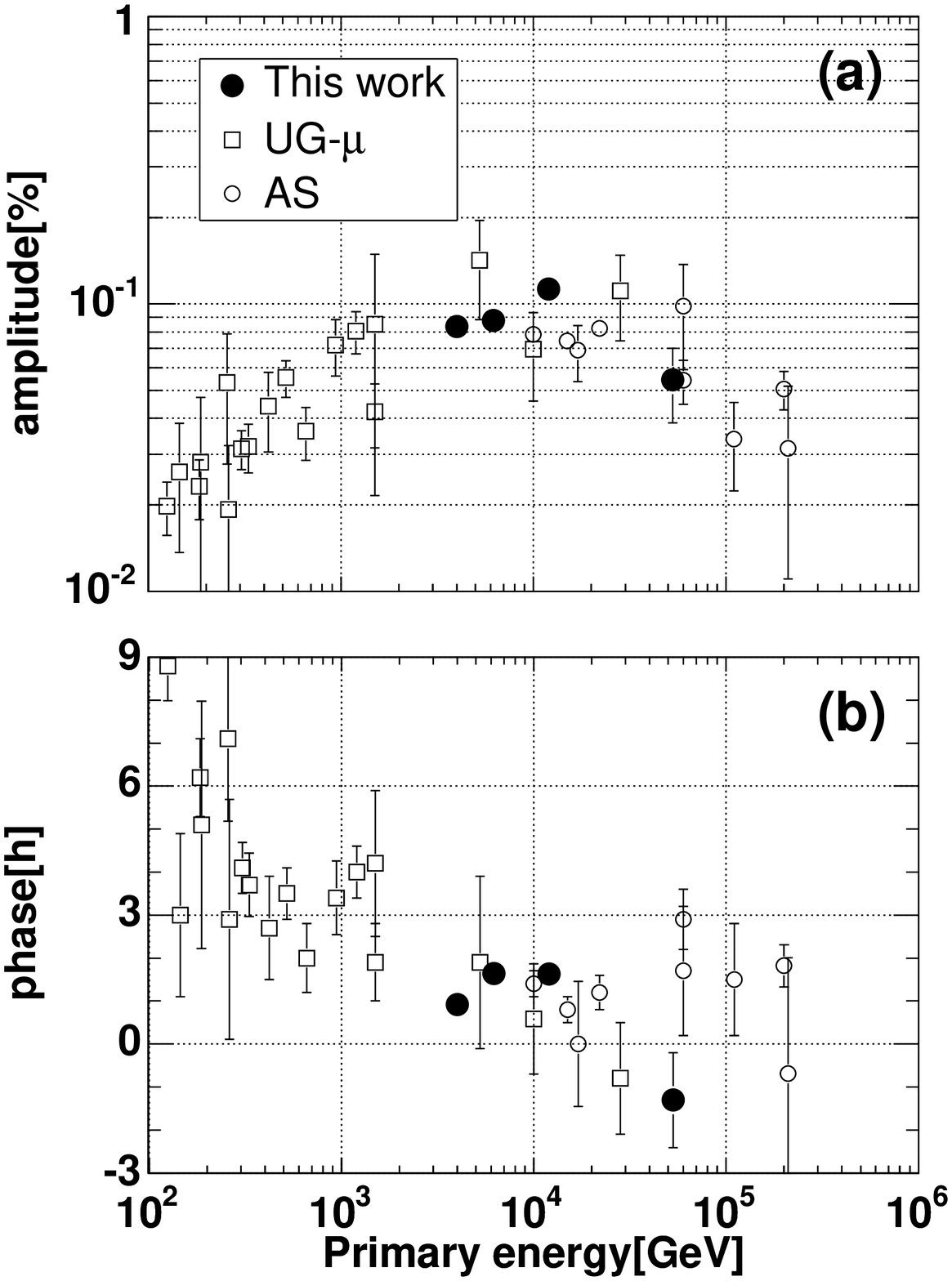}
\caption{The first harmonics of the sidereal daily variations obtained by underground
muon observations~\citep{Bercovitch,Thamby,nagashima-1985,swinson-1985,Andreyev,Lee,Ueno,cg1991,Fenton,
Mori,Munakata,mk-kamioka,MACRO2003} 
and by air-shower array experiments~\citep{Gombosi,Alexeenko,nagashima-1989,Bergamasco,EASTOP95,EASTOP,
nagashima-1998}. 
The amplitude (a) and the phase (b) of the first harmonics are plotted as a function of the primary 
cosmic-ray energy. 
Shown are Tibet III ({\it filled circles}), underground muon observations ({\it open squares}), 
and other air shower experiments ({\it open circles}). For clarity, the points are 
not labeled with citation information. The observed amplitude is divided by $\cos \delta$ 
for provisional correction of the difference in the representative declination ($\delta$). 
The data by Tibet III are averaged over all declinations. 
\label{ene_depend}
}
\end{figure}

On the basis of the SDV observed by the Norikura AS experiment, 
\citet{nagashima-1998} concluded that the amplitude of the ``Tail-In'' anisotropy 
with a maximum at $\sim6$ hours LST decreases with increasing primary energy above $\sim1$ TeV, 
while the ``Galactic'' anisotropy, with a minimum at $\sim12$ hours LST, remains constant. 
Furthermore, they suggested that the phase of the composite first harmonic vector turns 
counter-clockwise from $\sim3$ hours to $\sim0$ hours with increasing energy (the composite 
amplitude is also expected to decrease by $\sim30 \%$ when the amplitudes of two anisotropy 
components are equal at $\leq$ 1 TeV). 
The result from the present experiment, however, is apparently inconsistent with their model, 
as seen in Figure~\ref{sit-D_R_avedir}.
This figure showing the SDV averaged over all declinations ($-15^{\circ}$ to $75^{\circ}$), 
observed by the Tibet III, indicates that there is no significant energy dependence 
in the 24-hour profile of the SDV in this energy region. 
This is further confirmed in Figure~\ref{ene_depend} which depicts the energy dependence of 
the amplitude $A_{R}$ and the phase $\Phi_{R}$ of the first harmonic vector observed by the 
Tibet III, together with those from other experiments. 
It is readily seen that the first harmonic vector of the SDV is remarkably independent 
of the primary cosmic-ray energy in the multi-TeV region, contrary to the suggestion of NFJ. 
The amplitude and phase observed by the Tibet III are also summarized in Table~\ref{tab_values},
together with the $\chi^{2}$-values for the best-fitting of the first harmonic. 
The large $\chi^{2}$-values indicate that the first harmonic vector is not sufficient for a
precise representation of the full 24-hour profile.

\begin{figure}
\epsscale{0.6}
\plotone{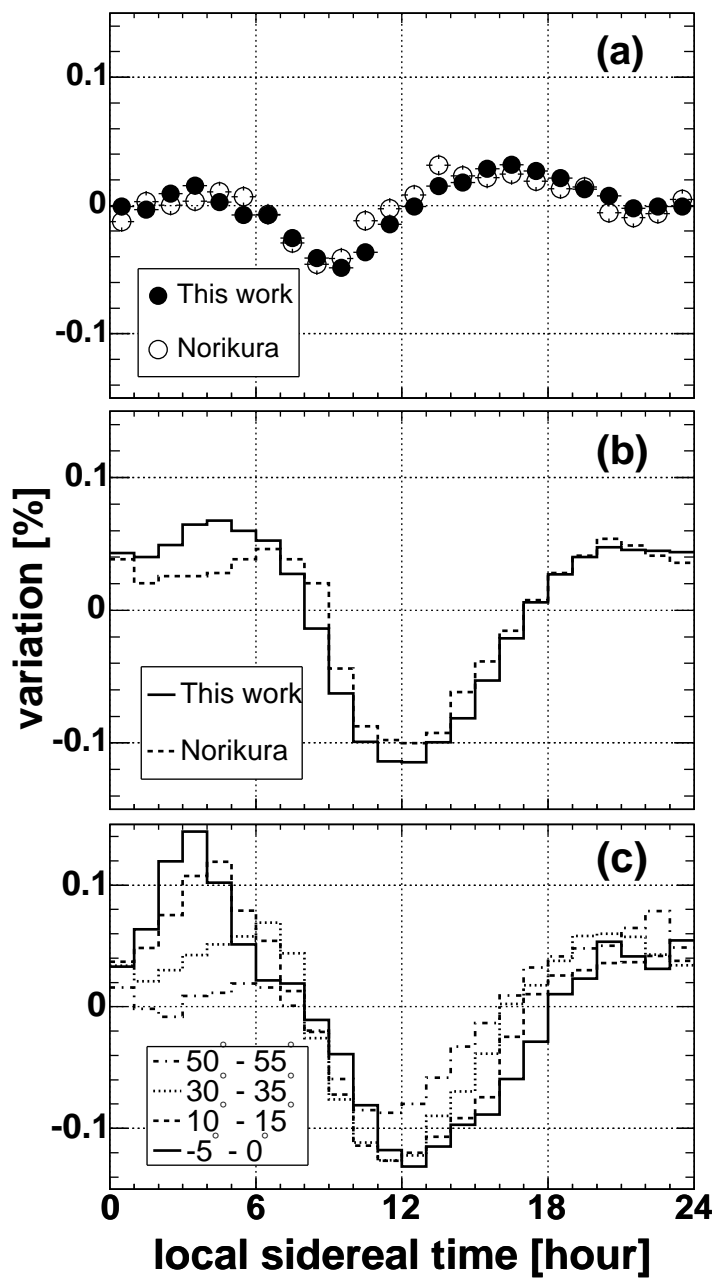}
\caption{The sidereal daily variations averaged over all primary energies. The
top (a) and middle (b) panels show respectively the differential variation $D(t)$
and the physical variation $R(t)$ averaged over all declinations, while the 
bottom panel (c) shows the physical variation $R(t)$'s in the four declination bands 
of $50^\circ-55^\circ$({\it dash-dotted line}), $30^\circ-35^\circ$({\it dotted line}), 
$10^\circ-15^\circ$({\it dashed line}), and $-5^\circ-0^\circ$({\it solid line}). 
The open circles in (a) and dashed histograms in (b) display the variations 
reported from the Norikura AS experiment~\citep{nagashima-1989}.
\label{each_decli_sftgt10}
}
\end{figure}

Figures~\ref{each_decli_sftgt10}(a) and~\ref{each_decli_sftgt10}(b) show the full 24-hour profile 
of the SDV averaged over all declinations and primary energies, together with the results of the
Norikura AS experiment(NFJ), which was not capable of resolving the declination of the incident 
direction. 
We find  that the Tibet III results are in a fairly good agreement with the Norikura data as far 
as the SDV averaged over all declinations is concerned. 
In Figures~\ref{each_decli_sftgt10}(b) displaying $R(t)$s, both results show the intensity deficit 
with a minimum around 12 hours, which is referred to as ``Galactic'' anisotropy by NFJ. 
The remarkable new features of the Tibet III measurement, however, become more apparent when one 
looks at the SDV in each declination band separately, as shown in Figure~\ref{each_decli_sftgt10}(c). 
It is evident that there is also an excess intensity with a maximum earlier than $\sim7$ hours LST. 
The phase of maximum changes from $\sim7$ hours to $\sim4$ hours LST, and the amplitude of the 
excess increases as the viewing direction moves from the northern hemisphere to the equatorial 
region.
This is qualitatively consistent with the anisotropy component first found in the sub-TeV region 
by underground muon detectors and referred to as ``Tail-In'' 
anisotropy~\citep{thn-1998,thn-1999,nagashima-1998}. 
The Tibet III experiment clearly shows that the ``Tail-In'' anisotropy continues to exist in the 
multi-TeV region covered by AS experiments. 
The present findings might require an alternative interpretation of the origin of this anisotropy,
since particle acceleration resulting in a $\sim0.1 \%$ anisotropy of the  multi-TeV cosmic rays
seems unlikely in the heliotail. 
Finally we note that the observations presented in this paper have been made possible by 
the high statistics and a good angular resolution provided by the Tibet III experiment. 
A more detailed study of the SDV based on the two-dimensional map is in progress. 
It seems also desireable to lower the energy threshold to the sub-TeV region, which would allow
the precise measurement of the heliospheric modulation of the sidereal anisotropy and may lead
to a better understanding of the large-scale magnetic structure of the heliosphere.  

\acknowledgments

This work is supported in part by Grants-in-Aid for Scientific Research on Priority Areas(712) (MEXT) 
and also for Scientific Research (JSPS) in Japan, and the Committee of the Natural Science Foundation 
and the Chinese Academy of Sciences in China. The authors thank J.Kota for a careful reading of this manuscript.

\clearpage
%
%
\begin{deluxetable}{crrr}
\tablecolumns{4} 
\tablewidth{0pt} 

\tablecaption{Amplitudes and phases of the first harmonic vector obtained with the Tibet III\label{tab_values}}

\tablehead{\colhead{Energy} & \colhead{Amplitude}           & \colhead{Phase} & \colhead{}\\ 
           \colhead{(TeV)}  & \colhead{($10^{-3}$ \%)} & \colhead{(h)}      & \colhead{$\chi^2$}}
\startdata 
4.0	& $83.4\pm5.2$ & $0.9\pm0.2$  & 338.1 \\ 
6.2	& $87.7\pm6.2$ & $1.6\pm0.3$  & 342.0 \\ 
12	& $112.6\pm6.7$& $1.6\pm0.2$  & 200.3 \\ 
53      & $54.4\pm15.8$& $-1.3\pm1.1$ &  58.2 \\
\enddata
\tablecomments{Amplitudes and phases observed by the Tibet III and plotted in Figure 2. The $\chi^{2}$
-values are deduced from harmonic analyses of the differential variations $D(t)$'s 
for the degree of freedom of 22.}

\end{deluxetable}

\clearpage

\clearpage

\clearpage

\end{document}